\documentclass[aps,twocolumn]{revtex4}

\usepackage{graphicx}
\usepackage{epsf}

\def\be#1{\begin{equation}\label{#1}}
\def\ee{\end{equation}}
\def\bea#1{\begin{eqnarray}\label{#1}}
\def\eea{\end{eqnarray}}
\def\sp{\hspace{.5em}}
\def\Eq#1{Eq.(\ref{#1})}
\def\Fig#1{Fig.(\ref{#1})}
\def\Figab#1#2{Fig.(\ref{#1}#2)}
\def\no{\nonumber \\}
\def\aM{a_M}
\def\adagM{a_M^{\dagger}}
\def\bI{b_{I}}
\def\bdagI{b^{\dagger}_{I}}
\def\bII{b_{II}}
\def\bdagII{b^{\dagger}_{II}}

\def\ketM#1{|#1\rangle_{M}}
\def\ketI#1{|#1\rangle_{I}}
\def\ketII#1{|#1\rangle_{II}}
\def\ketIII#1#2{|#1\rangle_{I}\otimes|#2\rangle_{II}}
\def\vacM{\ketM{0}}
\def\vacR{\ketIII{0}{0}}
\def\vacI{\ketI{0}}

\def\sOmega{\left( 1-e^{-2\pi\Omega} \right)^{-1/2}}
\def\expnOmega{e^{-2\pi\Omega n}}
\def\expmpiOmega{e^{-\pi\Omega}}
\def\expm2piOmega{e^{-2\pi\Omega}}
\def\Sum{\sum_{n=0}^{\infty}}

\def\sp{\hspace{.25em}}
\def\d#1#2{d^{\sp(#1)}_{\Omega,#2\vec{k}_\perp}}
\def\ddag#1#2{d^{\sp(#1)\dagger}_{\Omega,#2\vec{k}_\perp}}
\def\b#1#2{b^{\sp(#1)}_{\Omega,#2\vec{k}_\perp}}
\def\bdag#1#2{b^{\sp(#1)\dagger}_{\Omega,#2\vec{k}_\perp}}
\def\a#1{a_{#1\vec{k}_\perp,k^3}}
\def\adag#1{a^\dagger_{#1\vec{k}_\perp,k^3}}
\def\omegak{\omega_{\vec{k}}}
\def\kp{\vec{k}_\perp}
\def\bell#1{\ketM{\beta_{#1}}}

\begin{document}
\title{Teleportation with a uniformly accelerated partner}
\author{Paul M. Alsing}\email{alsing@hpcerc.unm.edu}
\affiliation{High Performance Computing, Education and Research Center and\\
Center for Advanced Studies, Department of Physics and Astronomy \\
University of New Mexico, Albuquerque, NM 87131}
\author{G. J. Milburn}\email{milburn@physics.uq.edu.au}
\affiliation{School of Physical Science,\\
University of Queensland, Brisbane, Australia}

\begin{abstract}
In this work, we give a description of the process of
teleportation  between Alice in an inertial frame, and Rob who is
in uniform acceleration with respect to Alice. The fidelity of the
teleportation is reduced  due to Unruh radiation in Rob's frame.
In so far as teleportation is a measure of entanglement, our
results suggest that quantum entanglement is degraded in non
inertial frames.
\end{abstract}

\date{\today}
\maketitle

\section{Introduction}
The large and rapidly growing field of quantum information
science  is a vindication of Landauer's insistence that we
recognize the physical basis of information storage, processing
and communication\cite{Landauer}.  Quantum information science is
based on the discovery that there are physical states of a quantum
system which enable tasks that cannot be accomplished in a
classical world. An important example of such a  task is quantum
teleportation\cite{Bennett-tele}.   Teleportation, like most
recent ideas in quantum information science, is based squarely on
the physical properties of non-relativistic quantum systems.

Recognizing that information science must be grounded in our
understanding of the physical world, one is prompted to ask how
relativistic considerations  might impact tasks that rely on
quantum entangled states. There has recently been some interest in
this question for inertial frames. While Lorentz transformations
cannot change the overall quantum entanglement of a bipartite
state\cite{Peres,Alsing},  they can change which properties of the
local systems are entangled. In particular, Gingrich and
Adami\cite{GA2002} showed that under a Lorentz
transformation the initial entanglement of just the spin
degrees of freedom of two spin half particles can
be transferred into an  entanglement  between both the spin and momentum
degrees of freedom.  Physically this means that detectors, which
respond only to spin degrees of freedom,  will see a reduction of
entanglement when they are moving at large uniform velocity.   Put
simply,  the nature of the entanglement resource depends on the
inertial reference frame of the detectors. A similar result  holds
for photons\cite{Gingrich03}

In this paper however, we wish to consider quantum entanglement  in
non-inertial frames. In order to make the discussion physically
relevant, we concentrate on a particular quantum information task;
quantum teleportation. We will show that the fidelity of
teleportation is compromised when  the  receiver is making
observations in a uniformly accelerated frame. This is quite
distinct from any reduction in fidelity through the Lorentz mixing
of degrees of freedom noted by Gingrich and Adami\cite{GA2002}.
Rather it is direct consequence of the existence of Unruh-Hawking
radiation for accelerated observers.  In so far as teleportation
fidelity is an operational measure of quantum entanglement, our
results suggest that quantum entanglement may not be preserved in
non-inertial frames.

\section{Uniformly Accelerated Observers}
\subsection{Preliminaries}
\begin{figure}[h]
\centering
\includegraphics[height=3.0in,width=2.0in,angle=-90]{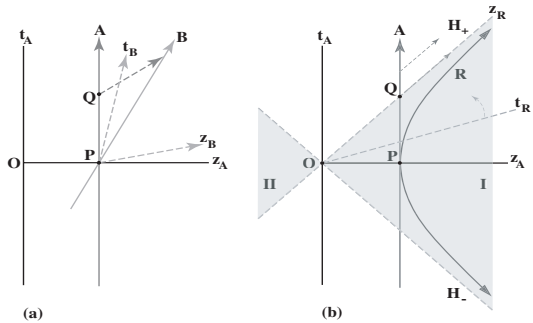}
\caption{(a) Minkowski diagram for the case of Alice (dark gray
arrow) stationary and Bob (light gray arrow) travelling at
constant velocity. Alice and Bob share an entangled Bell state
at the event $P$ (see text). Alice can complete the teleportation
protocol by sending classical signals to Bob at a representative
event $Q$. The entanglement fidelity of a state $\Phi$ is
unaltered if viewed from either Alice's or Bob's rest frame.
(b) Alice (dark gray arrow) is again stationary, while Rob (dark gray
hyperbola) undergoes constant acceleration. Alice and Rob share
an entangled Bell state at the common point $P$.
The light-like lines $\mathcal{H}_-$ and $\mathcal{H}_+$ form past and future particle horizon
corresponding to Rob's proper times $t_R = -\infty$ and $t_R =+\infty$ respectively.
At the event $Q$ Alice crosses $\mathcal{H}_+$ (in
her finite proper time $t_A$), and can no longer communicate with
Rob. Bob, however, can still send signals to Alice across $\mathcal{H}_+$.}\label{accelframe}
\end{figure}
Let Alice be an inertial Minkowski observer with zero velocity, located
at the point $P$ as shown in \Figab{accelframe}{a}.
Another inertial observer Bob is travelling with positive constant velocity $v < c$
in the $z$ direction with respect to Alice,
and their positions are coincident at the point $P$
whereupon they each share one part of an entangled Bell state.
The textbook teleportation protocol \cite{NC} proceeds as usual with Alice sending the
results of her measurement to Bob at the point $Q$, say by photons, so that Bob will
eventually receive them, and be able to rotate his half of the shared entangled state
into the state $\ketM{\psi} = \alpha\ketM{0} + \beta\ketM{1}$ that Alice wishes to teleport
(where the $M$ subscripts denotes a Minkowski state).

The situation is drastically different for the observer Rob who travels with constant acceleration $a$
in the $z$ direction with respect to Alice. Alice's and Rob's position coincide at the point
$P$ where again they instantaneously share an entangled Bell state, of which Rob takes one qubit on his journey.
In Minkowski coordinates Rob's world line takes the form
\be{1}
t_R(\tau) = a^{-1} \sinh{a\tau}, \qquad z_R(\tau) = a^{-1} \cosh{a\tau},
\ee
where $\tau$ is the proper time along the world line.
Rob's trajectory is a hyperbola in Minkowski space bounded by the  light-like asymptotes $\mathcal{H}_-$ and $\mathcal{H}_+$ which
represents Rob's past and future horizons with $\tau = -\infty$ and $\tau = \infty$ respectively.
The shaded region in the right half of the Minkowski plane in \Figab{accelframe}{b}
where Rob is constrained to move is
called the \textit{Right Rindler Wedge} (RRW) and is labelled with the roman numeral $I$. In general,
a point in the RRW can be labelled by the \textit{Rindler coordinates} $(\eta,\zeta)$
which are related to Minkowski coordinates $(t,z)$ by
\be{2}
t = \zeta \sinh{\eta}, \qquad z = \zeta \cosh{\eta},
\ee
where $-\infty < \eta< \infty$ and $0 < \zeta < \infty$. Lines of constant $\zeta$ are hyperbolas
within the RRW and lines of constant $\eta$ are straight lines through the origin. The past horizon
$\mathcal{H}_-$ corresponds to $\zeta = 0, \eta = -\infty$ while the future horizon $\mathcal{H}_+$ corresponds to
$\zeta = 0, \tau = \infty$.

With the same coordinate transformation given in \Eq{2}, the
region $-\infty < \zeta < 0$ and $-\infty < \eta < \infty$ is called the \textit{Left Rindler Wedge} (LRW) and
is labeled by the roman numeral $II$. In this region, the lines of constant $\eta$ run in the opposite
sense than in $I$. Region $I$ is causally disconnected from $II$ and no signal from one region can
propagate into the other region. The metric for Minkowski space is given by
\be{3}
ds^2 = dz^2 - dt^2 = d\zeta^2 - \zeta^2 d\eta^2.
\ee

It is well appreciated now \cite{fulling,unruh,BD,takagi,thooft,pringle} that the quantization
of fields in Minkowski and Rindler coordinates are inequivalent, implying that the
RRW vacuum seen by Rob $\vacI$ is different than the Minkowski vacuum seen by Alice $\vacM$.
The celebrated result of Davies and Unruh \cite{unruh} is that the Minkowski vacuum can be written in terms of
the region $I$ and $II$ states (for a scalar field) as
\be{4}
\vacM = \prod_{\Omega,\kp} \sOmega \Sum \expnOmega \ketIII{n_{\Omega,\kp}}{n_{\Omega,-\kp}},
\ee
where $\Omega \equiv \omega_R / (a/c)$ with  $\omega_R$ the frequency of a Rindler particle. The Minkowski vacuum as
given by \Eq{4} is a two-mode squeezed state \cite{WallsMilburn} which for each mode
$(\Omega,\kp)$ has the general form
\be{5}
\vacM \sim \frac{1}{\cosh r} \Sum \tanh^nr \, \ketIII{n}{n},
\ee
with
\be{5b}
\cosh r = \sOmega, \; \sinh r = \expmpiOmega \, \sOmega.
\ee
Note that $\vacM$ can be written as $S(r)\,\vacR$ where the two-mode squeezing
operator is given by
$S(r) \equiv \exp[r\,(b_I b_{II} - b^\dagger_I b^\dagger_{II})] $ \cite{WallsMilburn}.
The evolution of a Minkowski state vector is affected by the unitary operator
$e^{-i H_M t}$ where for a single mode (ignoring transverse momentum degrees
of freedom) $H_M = \hbar\omega_M \adagM \aM$.
For Rindler states the evolution proceeds via $e^{-i H_R \tau}$ where
\bea{6}
H_R &\equiv& H_I - H_{II} \\
\label{6b}
H_I &=& \hbar\omega_R \bdagI \bI, \qquad H_{II} =\hbar\omega_R \bdagII \bII.
\eea
The minus sign in \Eq{6b} stems from the sense of time essentially flowing
"backwards" in region $II$ (i.e. for $a<0$, $\eta(\tau) = a \,\tau$ is
a decreasing function of $\tau$).

For Rob, who lives in region $I$, all his observables can be
written solely in terms of $\bI$ and $\bdagI$ operators. Since he
is causally disconnected from region $II$, Rob must reduce any
density matrix describing both Rindler wedges to one appropriate
to region $I$ only, by tracing out over region $II$. Thus he
perceives the Minkowksi vacuum as a thermal state,
\be{7}
\rho^{(I)}_{\vacM} \equiv Tr(\vacM\langle 0|) = \left(
1-\expm2piOmega  \right) \Sum \expnOmega \ketI{n} \langle n |. \ee
The exponential terms can be written as $\exp(-\hbar\omega_R/k_B
T_U)$ with the \textit{Unruh temperature} $T_U$ is given by
(in units of $k_B=1$)
\be{8}
T_U \equiv \frac{\hbar a}{2\pi c} = \frac{\hbar}{2\pi c\,\zeta_0},
\ee
where $\zeta(\tau) = \zeta_0 = 1/a$ is the constant
Rindler position coordinate of Rob's stationary world line.

\subsection{Relationship between Minkowski and Rindler modes}
We will use two-photon states of the
electromagnetic field \cite{note1}
for the Bell state and so must consider  Fock states, other than
the vacuum state, for the Rindler observer. This is easily done by
a consideration of how the creation and anihilation operators
transform. The relationship between the Minkowski and Rindler
modes is given by the Bogoliubov transformation \be{9}
\b{\sigma}{} = \int d^3 k' \left( \alpha \alpha_{kk'}^{(\sigma)}
a_{k'} +
                                  \beta_{kk'}^{(\sigma)} a^\dagger_{k'}
                           \right)
\ee
where the notation of \cite{takagi} has been adopted, namely
 $\sigma = (+,-)$ refers to region $I$ and $II$ respectively, $k=(\Omega,\vec{k}_\perp)$ and
 $k' = (\vec{k}_\perp, k^3)$. Modes in Minkowski space are specified by the wave vector
 $\vec{k}\equiv (\vec{k}_\perp, k^3)$ where $\vec{k}_\perp = (k^1,k^2)$ are the components of
 the momentum perpendicular to the direction of Rob's acceleration. The Minkowski frequency is
 given by $\omegak = \sqrt{m^2 + \vec{k}^2}$.
 Modes in Rindler space
 are specified by a positive energy Rindler frequency $\Omega$ and $\kp$.
 The Bogoliubov transformation can be put into a more transparent form by introducing
 the \textit{Unruh modes} $\d{\sigma}{}$ and $\ddag{\sigma}{}$. The Unruh modes arise
 by considering the fourier transform of the usual Minkowski plane waves
 $[(2\pi)^3 \, 2\omegak]^{-1/2}\exp(\vec{k}\cdot\vec{x} - \omegak t)$ in terms of
 the Rindler proper time $\tau$ \cite{pringle}. These complete, orthonormal set of modes exits
 over all of Minkowski space and can be "patched together" to form the two complete orthonormal
 set of Rindler modes which have finite support in either region $I$ or region $II$.
 The physical significance of the Unruh modes is that they diagonalize the generator of
 Lorentz boosts \cite{takagi}, which in Minkowski coordinates is given by
 $$
    M^{\alpha\beta} = \int d^3x (x^{\alpha} T^{0\beta} -x^{\beta} T^{0\alpha} ).
 $$
 The restriction of the generator of boosts in the $z$ direction
 to region $I$ gives the Rindler Hamiltonian $H_R = \left. M^{03}\right|_I$.

The relationship between the Unruh modes and the Minkowski modes is given by \cite{takagi}
\be{10}
\a{} = \sum_\sigma \int_0^\infty d\Omega p_\Omega^{(\sigma)*}(k^3) \d{\sigma}{}.
\ee
which can be inverted to give
\be{11}
\d{\sigma}{} = \int_{-\infty}^\infty dk^3 p_\Omega^{(\sigma)}(k^3) \a{}.
\ee
In the above expression, the functions $p_\Omega^{(\sigma)}(k^3)$ form
a complete orthonormal set and are given by
\be{12}
p_\Omega^{(\sigma)}(k^3) = \frac{1}{(2\pi\omegak)^{1/2}}
\left( \frac{\omegak + k^3}{\omegak - k^3}  \right)^{i\sigma\Omega/2}
\ee
which are essentially phase factors.
Since by \Eq{11} the Unruh annihilation operator is a sum over only
Minkowski annihilation operators, it too annihilates the Minkowski vacuum.
\be{13}
\a{} \vacM = 0, \qquad \d{\sigma}{\pm} \vacM = 0.
\ee
Finally, the Unruh modes are related in a natural way to the Rindler modes
through the following Bogoliubov transformation
\be{14}
\left[
\begin{array}{c}
  \d{+}{} \\
  \ddag{-}{-}
\end{array}
\right]
=
\left[
\begin{array}{cc}
  \cosh r & -\sinh r \\
  -\sinh r & \cosh r
\end{array}
\right]
\,
\left[
\begin{array}{c}
  \b{+}{} \\
  \bdag{-}{-}
\end{array}
\right]
\ee
with the hyperbolic functions of $r$ related to the Rindler frequency $\Omega$ by \Eq{5b}.
The operators $b^{(+)}$ and $b^{(-)}$ annihilate the RRW vacuum $|0\rangle_+$ and LRW vacuum
$|0\rangle_-$ respectively, and commute with each other.

By \Eq{10} we see that a given Minkowski mode of frequency $\omegak$ is spread
over all positive Rindler frequencies $\Omega$, as a result of the
Fourier transform relationship between $\a{}$ and $\d{\sigma}{}$.
We now simplify our analysis by considering the effect of teleportation of
the state $\ketM{\psi} = \alpha \ketM{0} + \beta \ketM{1}$ by the Minkowski
observer Alice to a single Rindler mode of the RRW observer Rob. That is, we consider
only the mode $(\Omega,\kp)$ in region $I$ which is distinct from the mode $(\Omega,-\kp)$
in the same region. As such, we can consider only the $\sigma = (+)$ contribution
and drop the unessential phase factors in \Eq{10}. The single Rindler mode component of
the Minkowski vacuum state we are interested is then
\be{15}
\vacM \to \frac{1}{\cosh r} \Sum \tanh^nr \ketIII{n_{\Omega,\kp}}{n_{\Omega,-\kp}}.
\ee
The relevant Bogoliubov transformation can now be written as
\be{16}
\adag{} \to \ddag{+}{} = \cosh r \, \bdag{+}{} - \sinh r \, \b{-}{-}.
\ee
From here on we drop all the frequency and momentum subscripts and replace
the labels $\pm$ by $I$ and $II$, keeping
in mind the full definitions in \Eq{15} and \Eq{16}.

\section{Teleportation from a Minkowski observer to a Rindler observer}
Let us first begin by briefly recalling the usual teleportation protocol, between
Minkowski observers Alice and Bob [\Figab{accelframe}{a}], as given in \cite{NC}.
Our two qubit entangled state will be encoded as entangled Fock states of the electromagnetic field.
Alice wishes to teleport the state $\ketM{\psi} = \alpha \ketM{0} + \beta \ketM{1}$
to Bob. Let Alice and Bob share the entangled Bell state
$\bell{00} = 1/\sqrt{2}(\,\ketM{0}\otimes\ketM{0} + \ketM{1}\otimes\ketM{1}\,)$.
The input state to the system is then $\ketM{\Psi_0} = \ketM{\psi}\,\bell{00}$.
Alice performs a CNOT gate on $\ketM{\psi}$ and her portion of $\bell{00}$,
and then passes the first qubit of the output state through a Hadammard gate.
Upon making a joint projective measurement on her two qubits with the result
$\ketM{l}\otimes\ketM{m}$ with $l, m = \{0,1\}$, the full state is projected
into $\ketM{l}\otimes\ketM{m}\otimes\ketM{\phi_{l,m}}$ where Bob's state is
given by
$\ketM{\phi_{lm}}\equiv x_{lm}\ketM{0} + y_{lm}\ketM{1}$. Here we have defined
the four possible outcomes as
$(x_{00},y_{00}) = (\alpha,\beta)$,
$(x_{01},y_{01}) = (\beta,\alpha)$,
$(x_{10},y_{10}) = (\alpha,-\beta)$, and
$(x_{11},y_{11}) = (-\beta,\alpha)$.
After receiving the classical information $\{l,m\}$ of the result of Alice's
measurement, Bob can rotate his qubit of the entangled state into $\ketM{\psi}$
by applying the operations $Z^l_M\,X^m_M$ to $\ketM{\phi_{lm}}$.
The fidelity of the teleported state is unity in this idealized situation.

Alice now wishes to perform this same teleportation protocol with the
uniformly accelerated Rob. The calculation proceeds straightforwardly once
we remember two things. First, the states $\ketM{0}$ and $\ketM{1}= a^\dagger_M \ketM{0}$
of the initial state $\ketM{\Psi_0}$ associated with Rob must be expanded in terms of
the Rindler states $\ketI{n}\otimes\ketII{m}$. This can be accomplished by using the
single mode Minkowski vacuum as given in \Eq{15} and the Bogoliubov transformation \Eq{16}.
Here we are imagining the situation where Rob has instantaneously accelerated
to the value $a$ at $\tau=0$ at the point $P$ in \Figab{accelframe}{b}
where Alice and Rob initially share the entangle state $\bell{00}$.
Second, since the result of Rob's acceleration inextricably
creates a particle horizon which keeps him causally disconnected from
region $II$, his final state is produced by tracing out
over region $II$. Thus, since any Minkowski Fock state is a correlated state
of region $I$ and $II$ Fock states, there is no hope of completely
teleporting the state $\ketM{\psi} = \alpha \ketM{0} + \beta \ketM{1}$
in the presence of this partial trace operation by Rob. However, the best state
we might expect Rob to obtain as an end product of the teleportation protocol
would be the region $I$ analogous version of
the Minkowski transported state $\ketM{\psi}$, namely
\be{17}
\ketI{\psi} = \alpha \ketI{0} + \beta \ketI{1}.
\ee
We might call this a \textit{thermally teleported state} since
Rob perceives all Minkowski states through the haze of the thermal
vacuum that he moves through.
If $\rho^{(I)}_{lm}$ is Rob's density matrix after Alice performs the
operations on her qubits, then we can measure the fidelity of the
thermally teleported state as
\be{18}
F^{(I)} \equiv Tr_I\Big(\ketI{\psi}\langle\psi|\,\rho^{(I)}\Big) = \sp_I\langle\psi|\rho^{(I)}\ketI{\psi}.
\ee

Using \Eq{5} for $\vacM$ and \Eq{16} for $a^\dagger_M$ we find
\be{19}
\ketM{1} = \frac{1}{\cosh^2 r} \Sum \tanh^nr \, \sqrt{n+1}\,\ketIII{n+1}{n}.
\ee
When Alice sends the result of her measurement $\{l,m\}$, which can
be received by Rob, if Alice has not yet crossed Rob's future horizon $\mathcal{H}_+$,
Rob's state will be projected into
\bea{20}
\rho^{(I)}_{lm} &\equiv& \Sum \sp _{II}\langle n\ketM{\phi_{lm}} \langle \phi_{lm}\ketII{n} \no
 &=& \Sum \frac{\tanh^{2n} r}{\cosh^2 r}
 \left[ \left( |x_{lm}|^2 + n \frac{|y_{lm}|^2}{\sinh^2 r}\right) \ketI{n}\langle n| + \right.\no
 &+& \left. \frac{x_{lm} y^*_{lm}}{\cosh r} \,\sqrt{n+1}\,\ketI{n+1}\langle n| \right.\no
 &+&
\left. \frac{x^*_{lm} y_{lm}}{\cosh r} \,\sqrt{n+1}\,\ketI{n}\langle n+1| \right].
\eea
Note that as the acceleration becomes large i.e.  $r\to\infty$, Rob's state
is driven into the thermal vacuum state $\rho^{(I)}_{\vacM}$ of \Eq{7} and
all information has been lost due to thermalization.

Let us now compute the fidelity in \Eq{18} with the state
$\ketI{\phi_{lm}} =  x_{lm} \ketI{0} + y_{lm} \ketI{1}$ with $|x_{lm}|^2 + |y_{lm}|^2 = 1$,
which is the penultimate
state before Rob would perform the appropriate rotation to attempt to transform his
half of the (accelerated) entangled Bell state to its final form $\ketI{\psi}$.
We obtain
\bea{21}
F^{(I)}_{lm} &=& \sp_I\langle\phi_{lm}|\rho^{(I)}_{lm}\ketI{\phi_{lm}} \no
 &=& \frac{1}{\cosh^2 r} \,
 \left[
|x_{lm}|^4 + \left( \tanh^2 r |x_{lm}|^2 + \frac{|y_{lm}|^2}{\cosh^2 r}\right)\,|y_{lm}|^2 \right. \no
&+& 2 \left. \frac{|x_{lm}|^2 |y_{lm}|^2}{\cosh r}
 \right],
\eea
which is essentially a projection onto the $\{\ketI{0},\ketI{1}\}$  subspace
of $\rho^{(I)}_{lm}$.

\begin{figure}[h]
\centering
\includegraphics[height=3.0in,width=3.0in]{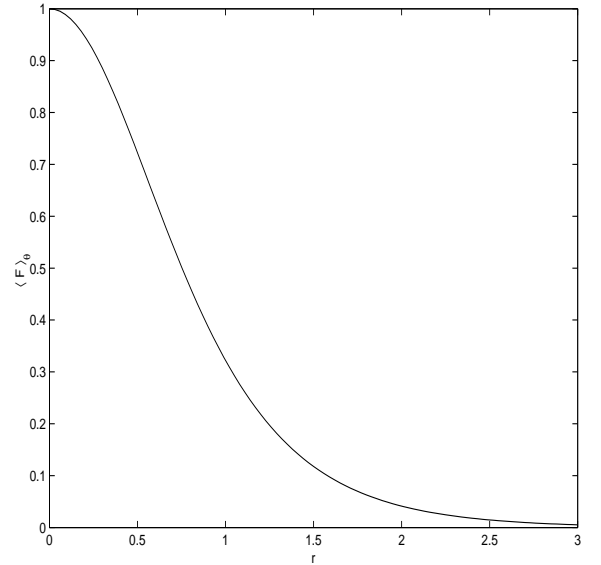}
\caption{Averaged Fidelity $1/\pi \int_0^\pi d\theta F^{(I)}_{lm}(r,\theta)$ over all possible
input states for $x_{lm} =\cos\theta$, $y_{lm} =\sin\theta$. Note that $r$ scales directly with the
acceleration $a$, so that $r=0$ corresponds to $a=0$.}\label{Fidr}
\end{figure}
In \Fig{Fidr} we plot $F^{(I)}_{lm}$ averaged over all possible input
states using the parameterization $x_{lm} =\cos\theta$ and $y_{lm} =\sin\theta$.
At $r=0$, corresponding to $a=0$, we
are back to the case of teleportation between Alice and Bob in Minkowski space,
and the fidelity is unity. Using the definition of $\Omega$ and $r$ in
\Eq{5b} we find
\bea{22}
\tanh r &=&  \exp[-\pi\omega_R / (a / c)] \quad \textrm{or} \no
r &\approx& \exp[-\pi\omega_R / (a / c)] \quad \textrm{for} \quad r\to 0.
\eea
For terrestrial experiments $a=g\sim 10 \, m/s^2$ and $a/c \sim 10^{-8} \,s^{-1}$
is such a small frequency, that for all frequencies $\omega_R$ of physical interest
the fidelity is near unity with incredible precision, (for $r=10^{-3}$ the fidelity
is is unity to within one part in $10^6$, and this still corresponds to
unphysically large accelerations).
Near the event horizon of a black hole appreciable accelerations can be obtained
such that the reduction of the fidelity from unity could be observed.
In that case an analogous teleportation scheme could be defined with Rob, stationary,
outside the event horizon and Alice freely falling into the hole.

In the above we have ignored the time evolution of the states, namely that
Alice's qubits evolve according to $e^{-i H_M t}$ while Rob's
qubits evolve via $e^{-i H_R \tau}$, as discussed in the previous section.
A simple calculation reveals that only change from the above
analysis is that $y_{lm}\to y_{lm} e^{-i\Omega\tau}$ in \Eq{19}.
However, since the fidelity in \Eq{21} depends only upon $|y_{lm}|^2$,
this phase factor does not contribute to the final result. From Rob's
point of view, he is able to receive the result of Alice's measurement
for all of his eternity, i.e $0 < \tau < \infty$. From Alice's point
of view, she cross Rob's future horizon $\mathcal{H}_+$ in a finite time $t=c/a$, after
which even light signals will not reach Rob (see point $Q$ in \Figab{accelframe}{b}).
For her, the teleportation protocol stops since she can no longer communicate
the results of her measurements to Rob. However, for all of her eternity, i.e.
$0 < t < \infty$ she can still receive persistent requests from Rob to
send the information. $\mathcal{H}_+$ is analogous to the "one-way membrane"
of an black hole event horizon, which Rob has "fallen" through.

\begin{figure}[h]
\centering
\includegraphics[height=3.0in,width=3.0in]{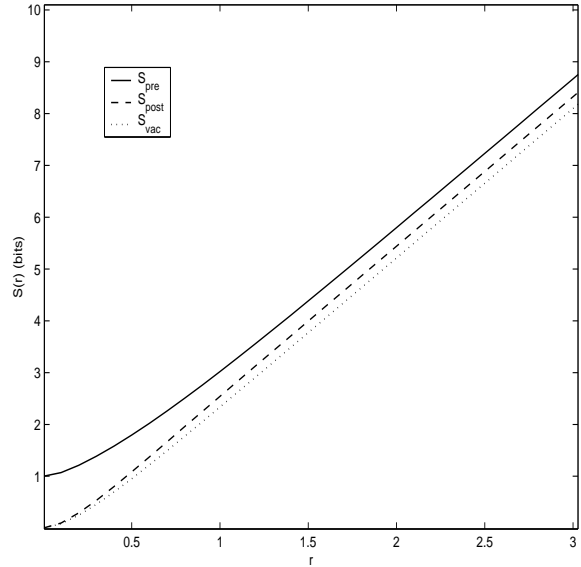}
\caption{Von Neumann entropies of Rob's pre-measurement state $S_{pre}$ (solid),
post-measurement state $S_{post}$ (dashed), and the vacuum $S_{vac}$ (dotted).
}
\label{Svsr}
\end{figure}

It is of some interest to consider the reduction of fidelity  in terms of entropy.
In \Fig{Svsr} we plot the von Neumann entropy $S = - Tr(\rho \log \rho)$
of Rob's pre-measurement state, post-measurement state upon learning the result
of Alice's measurement, and the vacuum state, as a function of $r$  (measured in bits).
The pre-measurement state is obtained from \Eq{20} by summing $(l,m)$ over all four possible input states,
which reduces it to a diagonal density matrix.
The post-measurement state is given by \Eq{20} with the input state to the teleportation
protocol chosen to be $\ketM{\psi} = 1/\sqrt{2} \, (\ketM{0} + \ketM{1})$.

For $r=0$, corresponding to $a=0$, Rob's pre-measurement state is just $\frac{1}{2} \mathbf{I}_{2\times 2}$,
half the 2-by-2 unit matrix in the $\{\ketM{0}, \ketM{1} \}$ sector (since $\ketI{n}\equiv\ketM{n}$
in this limit) and zero elsewhere, while the vacuum is just the normal Minkowski vacuum.
Note that $S_{pre}(0) = 1$  expressing Rob's (or here Bob's at $r=0$)
complete ignorance of the outcome of Alices's measurement. As $r$ increases all entropies
increase since the thermalization eventually drives all states to the vacuum (in the limit $r\to\infty$)
with equally weighted states,  where all information is lost. The energy for this
thermalization, of course, comes from the work supplied by the agent causing Rob's acceleration.

\begin{figure}[h]
\centering
\includegraphics[height=3.0in,width=3.0in]{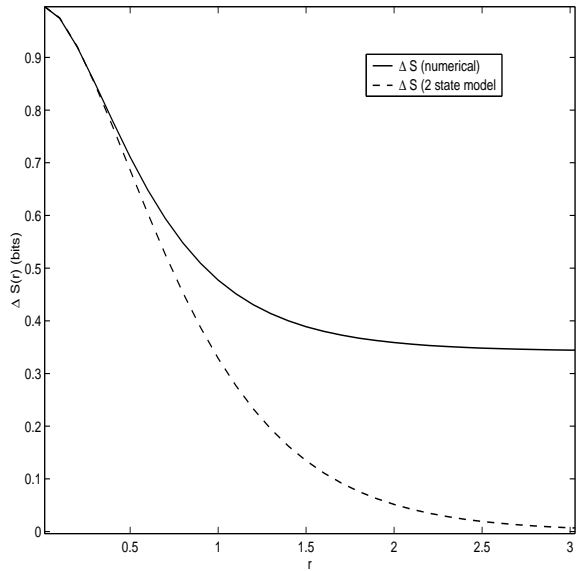}
\caption{Rob's entropy information gain (in bits) $\Delta S_{gain}
= S_{pre} - S_{post}$ upon receiving the Alice's measurement
results: numerical (solid) and $\Delta S_{gain}^{TSM}$ (dashed) for a 2-state
model using the $\{\ketI{0}, \ketI{1} \}$ sector from Rob's post
measurement state, with $x_{lm} = y_{lm} = 1/\sqrt{2}$. }
\label{dSvsr}
\end{figure}

Also plotted in \Fig{Svsr} is the entropy for Rob's post-measurement state $S_{post}$.
At $r=0$ Rob (or Bob) would absolutely know the state that was teleported to him and thus gain 1 bit
of information (the distance between the solid and dashed curve). However, as the
acceleration increases, this information gain decreases from unity to zero
as the Unruh temperature increases. This is seen as the solid curve in \Fig{dSvsr} which
plots $\Delta S_{gain} \equiv S_{pre}-S_{post}$.

Though $\Delta S_{gain}\equiv S_{pre} - S_{post}$ appears to be levelling out for large $r$
in \Fig{dSvsr}, it is doing so only very slowly.
Since, as mentioned above, all states are driven to the
infinite temperature vacuum as $r\to\infty$, these curves must eventually
merge. It is curious that they do so slowly. The dashed curve in
\Fig{dSvsr} is a 2-state model which uses only the (normalized) $\{\ketI{0}, \ketI{1} \}$ sector
of Rob's post-measurement state.
The entropies for this two-state model  and the vacuum can
then be computed analytically and the resulting entropy difference $\Delta S_{gain}^{TSM}$
agrees well with the numerical calculation $\Delta S_{gain}$ for small values
of $r\leq 1/2$. This two state model thermalizes much faster than the
full post-measurement state and so approaches zero more rapidly due the
finite number of states (2) used.
The interesting point to note is that even after Rob receives Alice's
classical information about the result of her measurement
\textit{Rob is prevented from regaining the full 1 bit of information if his
acceleration is non zero}. This is another way to see that from Rob's perspective
information appears to be lost.

\section{Discussion and Conclusions}

The main issues of teleportation between an inertial Minkowksi observer Alice
and a non-inertial, uniformly accelerated Rindler observer Rob are two fold.
First, as a result of the acceleration, the Minkowski vacuum that Rob moves
through (for a single Rindler mode $(\Omega,\kp)$) can be written as a
two-mode squeezed state with the component Fock states existing in causally
separated regions $I$ and $II$. Second, as a result of this fact, Rob's perceives the
Minkowski vacuum as a pure thermal state of temperature
$T_U$ as the inevitable result of his complete ignorance of region $II$.

In an attempt to teleport a state $\ketM{\psi} = \alpha \ketM{0} + \beta \ketM{1}$
to Rob, the best we can expect Rob to recover at the end of the protocol is
$\ketI{\psi} = \alpha \ketI{0} + \beta \ketI{1}$. In this work we have calculated
the fidelity of the state Rob receives with this best possible result $\ketI{\psi}$.
We have demonstrated how the fidelity decreases with increasing acceleration until
at high temperatures all information is lost and Rob perceives only the thermalized
vacuum state.

The model investigated here is equivalent to teleportation through two channels,
one of which is free space for Alice and the second which involves parametric
down conversion with the following caveat described below. In the second channel, a signal
mode $I$ and an idler mode $II$ experience a squeezing Bogoliubov transformation analogous to
\Eq{14} \cite{yurke}. Here $r$ is proportional to the coupling strength between
the signal and idler mode times the length of the crystal through which
the parametric down conversion takes place; higher interaction strengths and/or
longer interaction lengths corresponds to a higher Unruh temperature.
The caveat is that Rob, acting as say the signal mode, has no access
to information about the idler and therefore must trace out this information.
Performing the teleportation protocol in such a system is exactly analogous to
teleportation between a Minkowski and Rindler observer as considered in this work.
In the parametric down conversion model, Rob can choose to ignore the idler
information thus mimicking a Rindler observer. However, for an
accelerated observer, the existence of the horizons $\mathcal{H}_\pm$ is of fundamental
importance. Since region $I$ and $II$ are causally disconnected, there is
no way, even in principle, for Rob to have any information about region $II$,
and thus his state is always a reduced density matrix appropriate for region $I$.

We have given an explanation of the reduction of teleportation
fidelity in terms of the Unruh radiation seen by Rob in his frame.
Note that this is an operationally meaningful statement as Rob can
attempt to verify that he has not received the desired state
$\big(x_{lm}|0\rangle_I+y_{lm}|1\rangle_I\big)$ by local verification
measurements (e.g. a single photon interference experiment), and then
send the results to Alice. It would be quite easy  to arrange a
situation whereby Alice could tell unambiguously that Rob had
received the wrong state.  From an operational point of view Alice
would conclude that the shared entangled  resource  has become
decohered. It is well know that entanglement is a fragile resource
in the presence of environmental decoherence. It appears also to be a fragile
resource when one of the entangled parties undergoes acceleration.
While the degree of decoherence  is exceedingly small for practical
accelerations, the apparent connection between space time geometry
and quantum entanglement is intriguing.

\textit{Added Note} During the preparation of this work, the authors
became aware of the recent paper by Anderson \textit{et al} \cite{vanEnk} which also
discusses teleportation and the Unruh vacuum. However, that work
considers a physically different situation than the one presented here.
The authors use the \textit{mirror modes} of Audretsch and M\"{u}ller \cite{audretsch}
and consequently have the accelerated observers travelling on oppositely
directed hyperbolas, with Alice in region $I$ and Bob in the causally
disconnected region $II$. The teleportation protocol is then interpreted
from the point of view of a Minkowski observer Mork. In this work, we
consider a setup between observers, one stationary, the other accelerated, who remain
causally connected to each other during the teleportation protocol.

\acknowledgements{The authors wish to thank Jon P. Dowling for stimulating discussions
on this and other topics in relativistic quantum information theory.}



\begin{thebibliography}{99}
\bibitem{note1} The quantization of field modes discussed this work
has actually been carried out for scalar particles. However, the main results
of this work are not modified by the inclusion of spin or helicity degrees of freedom.
\end{thebibliography}
\end{document}